\documentclass{ws-ijmpa}

\begin{document}

\markboth{P.~Adlarson, M.~Zieli\'{n}ski}
{Measurement of the $\eta\rightarrow\pi^+\pi^-\pi^0$ decay with WASA-at-COSY detector}

%
\catchline{}{}{}{}{}
%

\title{Measurement of the $\eta\rightarrow\pi^+\pi^-\pi^0$ decay with WASA-at-COSY detector}

\author{PATRIK ADLARSON$^1$, MARCIN ZIELI\'{N}SKI$^2$\footnote{emial: patrik.adlarson@fysast.uu.se, m.zielinski@uj.edu.pl}~~for the WASA-at-COSY Collaboration}
\address{$^1$Institute for Physics and Astronomy, Uppsala University, Uppsala, 751 20, Sweden\\
               $^2$Institute of Physics, Jagiellonian University Krak\'{o}w, 30-059 Krak\'{o}w, Poland}
\maketitle


\begin{abstract}
One of the objectives of the physics programme of the WASA-at-COSY facility is to study the isospin violating $\eta$ hadronic decays into
$\pi^+\pi^-\pi^0$ systems driven by the term of QCD Lagrangian which depends on the d and u quark mass difference.
These studies can be made in terms of the Dalitz plot parameters describing the density population which is proportional to the square of the amplitude $\left|A(x,y)\right|^2$.
This contribution describes the current status of the analysis  of the $\eta\rightarrow\pi^+\pi^-\pi^0$ decay  in the $pd\rightarrow  ^{3}He \eta$ 
and as well in the $pp\to pp\eta$ reaction with WASA-at-COSY.

\end{abstract}

\ccode{PACS numbers: 11.25.Hf, 123.1K}

\section{Motivation}
The isospin violating strong decay $\eta\rightarrow\pi^+\pi^-\pi^0$ allows access to light quark mass ratios. 
At lowest order of chiral perturbation theory (ChPT) the amplitude is proportional to the light quark mass difference $(m_d-m_u)$ and may be written as 
\begin{equation}
A \propto \frac{m_d-m_u}{F_{\pi}^2}\left( 1 + \frac{3(s-s_0)}{m_{\eta}^2-m_{\pi}^2} \right),
\end{equation} 
where $F_{\pi}$ is the pion decay constant, $s=(p_{\pi^+}+p_{\pi^-})^2=(p_{\eta}-p_{\pi^0})^2$ and $s_0=\frac{1}{3}(m_{\eta}^2+2m_{\pi^+}^2+m_{\pi^0}^2)$.
At higher order of ChPT it has been found that final state pion interaction contribute to the decay width\cite{CR81}$^{,}$\cite{JG85}. The decay width scales as
$\Gamma=\left(\frac{Q_D}{Q}\right)^4 \bar{\Gamma}$,
where $Q^2=\frac{m_s^2-\hat{m}^2}{m_d^2-m_u^2}$, $\hat{m}=\frac{1}{2}(m_u+m_d)$,
and the decay width $\bar{\Gamma}$ and $Q_D=24.2$ are calculated in the Dashen limit\cite{RD69}. This scaling works under the pre-requisite that $\bar{\Gamma}$ is understood reliably. 
To test this, theoretical predictions and experimental measurements of pion kinematical distributions 
may be compared in a Dalitz plot, where the axes are defined as
$x=\sqrt{3}\frac{T_+-T_-}{Q_{\eta}}$, $y=\frac{3T_0}{Q_{\eta}}-1$.
\noindent Here  $T_+$, $T_-$ and $T_0$ denote the kinetic energies of $\pi^+$, $\pi^-$ and $\pi^0$ in the rest frame of the $\eta$ meson, 
and $Q_{\eta}=T_++T_-+T_0=m_{\eta}-2m_{\pi^+}-m_{\pi^0}$.
The standard way to parametrize the Dalitz plot density is a polynomial expansion around the center point: 
$\left|A(x,y)\right|^2 \propto 1+ay+by^2+dx^2+fy^3+gx^2y+...$
where $a, b, ...,g$ are the Dalitz plot parameters. 
The experimental results are dominated by KLOE with a Dalitz plot containing 1.34$\cdot$10$^6$ events \cite{FA08}.
This result shows a significant deviation of parameters $b$ and $f$ in comparison to the theoretical predictions based on ChPT. 
It is therefore important to perform an independent measurement, which is one of the aims of the WASA-at-COSY experiment. 

\section{$pd\rightarrow  ^{3}He \eta$ measurement}
\begin{figure}[t!]
\begin{center}
\parbox[c]{0.30\textwidth}{
\centering
\includegraphics[width=0.35\textwidth]{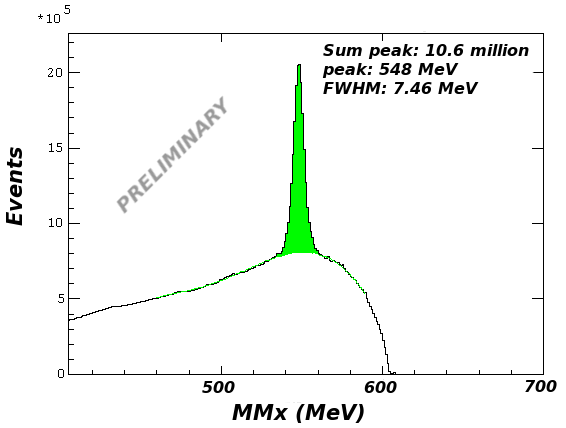}
}
\parbox[c]{0.5\textwidth}{
\centering
\includegraphics[width=0.35\textwidth]{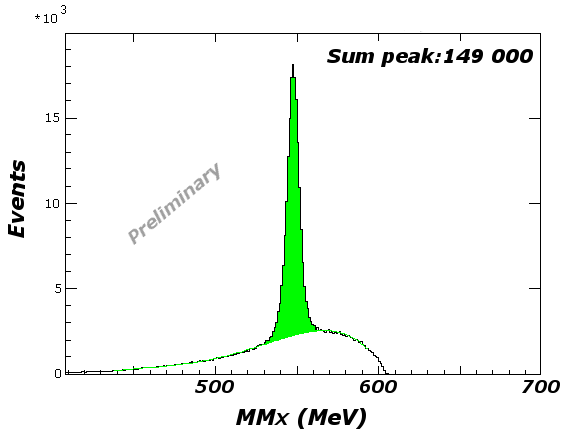}
}
\caption{({\bf left}) Missing mass for the 2008 data calculated from the identified $^{3}He$. 
({\bf right})  Missing mass after selecting 3$\pi$ candidates and including a cut on $MM(^{3}He \pi^+\pi^-)$ and $MM(^{3}He \pi^0)$.}
\label{MM(He)}
\end{center}
\end{figure}

\begin{figure}[h!]
\begin{center}
\parbox[c]{0.8\textwidth}{
\centering
\includegraphics[width=0.80\textwidth]{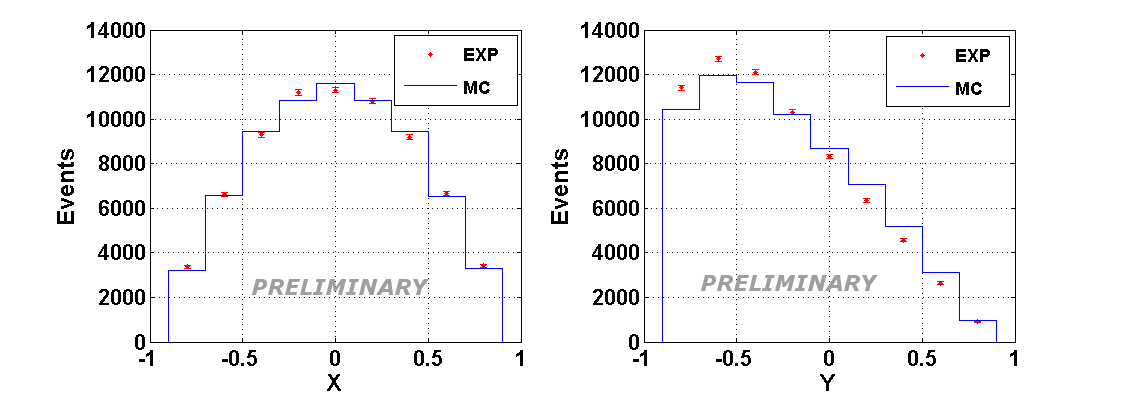}
}
\caption{Projections of Dalitz Plot, not corrected for acceptance and normalized to sum of experimental data: 
{\bf(left)} X-projection {\bf(right)} Y-projection. Solid line indicates MC data and points with error bars experimental values.}
\label{XYprojection}
\end{center}
\end{figure}

\noindent In 2008 and 2009 WASA-at-COSY\cite{AA04} measured $pd\rightarrow ^{3}He X$ reaction at beam energy 1~GeV, collecting $10^7$ and $2\cdot10^7$ $\eta$ mesons respectively. 
The missing mass with respect to  $^{3}He$ is used to tag the $\eta$ meson (Fig.~\ref{MM(He)} left). 
In addition two tracks of opposite charge are required in the Mini Drift Chamber in the angular range $30.5^{\circ}<\theta<150^{\circ}$. 
Furthermore two photons with an invariant mass close to $\pi^0$ are required. The $pd\rightarrow ^{3}He \pi\pi$ reaction is reduced by 
imposing conditions on the missing mass calculated for $^{3}He \pi^+\pi^-$ and the missing mass calculated for $^{3}He \pi^0$. 
The preliminary analysis yields 149 000 $\eta\rightarrow\pi^+\pi^-\pi^0$ candidates from the 2008 data, shown in Fig.~\ref{MM(He)}~right.
\noindent The experimental resolution is better for the $\eta$ four-momenta from $^{3}He$ compared to the information derived from the $\eta$ decay products. 
Therefore a kinematical fit for the reaction $pd\rightarrow ^{3}He\pi^+\pi^-\pi^0$ has been used with ${^3}He$ observables fixed and a cut on the 1\% level of the probability density function. 
To estimate the $\eta$ content in each Dalitz plot bin, a four-degree polynomial fit is performed over the background region and the fitted polynomial
 is subtracted in the signal region. The preliminary experimental results for the $x$,$y$ projections of the Dalitz plot are compared in Fig.~\ref{XYprojection} to Monte Carlo simulations of the $\eta\rightarrow\pi^+\pi^-\pi^0$ weighted with the tree-level prediction (equation 1).
\section{$pp\rightarrow pp\eta$ measurement}
\begin{figure}[h!]
\begin{center}
\parbox[c]{0.27\textwidth}{
\centering
\includegraphics[width=0.27\textwidth]{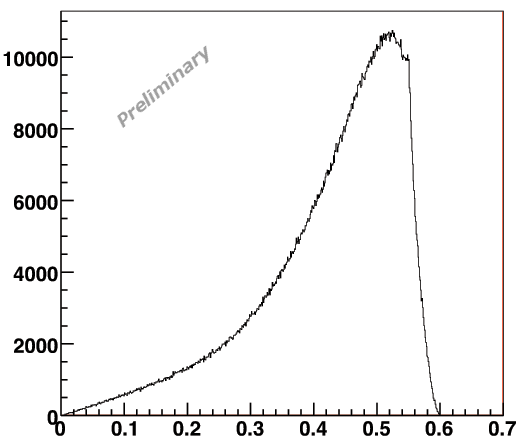}
}
\parbox[c]{0.3\textwidth}{
\centering
\includegraphics[width=0.27\textwidth]{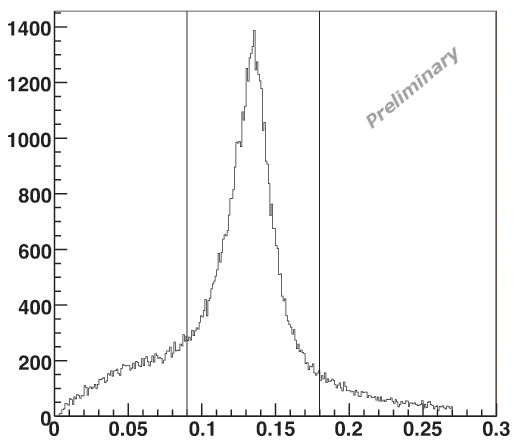}
}
\parbox[c]{0.3\textwidth}{
\centering
\includegraphics[width=0.27\textwidth]{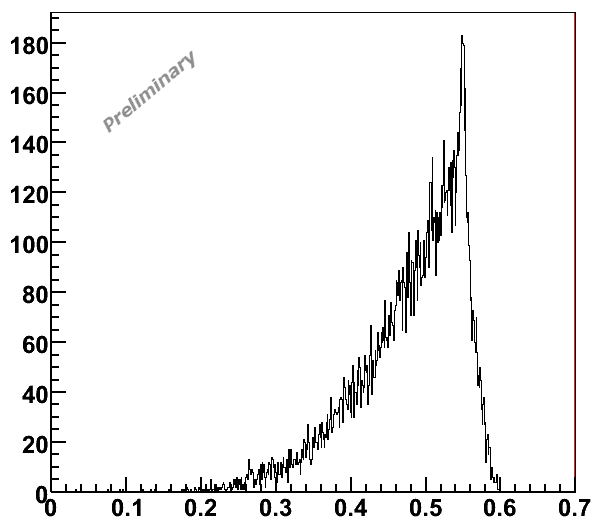}
}
\caption{{\bf{(left)}} Missing mass of the $pp\to ppX$ reaction. {\bf{(middle)}} Invariant mass ot two $\gamma$ with cut lines. 
{\bf{(right)}} Missing mass of two protons with the requirement of two $\gamma$ in coincidence.}
\end{center}
\label{mz}
\end{figure}

The measurement of the  $pp\to ppX$ reaction was conducted in 2008 and in 2010 at beam kinetic energy  1.4~GeV.
The collected sample of data yields about 10$^{8}$ produced $\eta$ mesons.  Two protons were registered
using plastic scintillator detectors and straw tube trackers and the charged pions were detected with plastic 
scintillator detectors and the Mini Drift Chamber. Two protons were used to tag the $\eta$ meson in the missing mass plot showed in Fig.~\ref{mz}~left (here we 
present data only from one run). 
The two $\gamma$ originating from the $\pi^0$ meson decay were registered in the electromagnetic calorimeter. The invariant mass of these$\gamma$ is required to be close to the mass of the 
$\pi^0$ (Fig.~\ref{mz}~middle). 
Requiring two $\gamma$ in coincidence with the two protons gives the missing mass as shown in Fig.~\ref{mz}~right.
\section{Outlook}
The work for both \textit{pd} and \textit{pp} data will be continued in order to obtain two independent determinations of the Dalitz plot density for the $\eta\rightarrow\pi^+\pi^-\pi^0$. This includes estimating systematical errors as well as tuning Monte Carlo simulation. 


\begin{thebibliography}{99}
\bibitem{CR81} C. Roiesnel, T. Truong, \textit{Nucl Phys B} \textbf{187} (1981) 293.
\bibitem{JG85} J. Gasser, H. Leutwyler,  \textit{Nucl Phys B} \textbf{250} (1985) 539.
\bibitem{RD69} R. Dashen, \textit{Phys Rev} \textbf{183} (1969) 1245.
\bibitem{FA08} KLOE Collab. (F. Ambrosini \textit{et al}.), \textit{JHEP} \textbf{05} (2008) 006, arXiv:0801.2642v2 [hep-ex].
\bibitem{AA04} WASA-at-COSY Collaboration, arXiv:nucl-ex/0411038, (2004).
\end{thebibliography}
\end{document}